\documentclass[a4paper]{article}

\usepackage{fullpage} 
\usepackage{pstricks}
\usepackage{multicol}
\usepackage{graphicx}
\usepackage{movie15}
\usepackage{hyperref}
\usepackage[version=3]{mhchem}
\usepackage{amsmath} 
\usepackage{amssymb} 
\usepackage{amsfonts}
\usepackage{color}
\usepackage{soul}
\usepackage[margin=1.7cm,top=3cm,bottom=2.8cm,centering]{geometry}

\definecolor{rosepale}{rgb}{1.0, 0.7, 1.0}

\newcommand{\be}{\begin{equation}}
\newcommand{\ee}{\end{equation}}
\newcommand{\bea}{\begin{eqnarray}}
\newcommand{\eea}{\end{eqnarray}}

\def\li{{\rm Li}}
\def\tli{\L{\rm i}}
\renewcommand{\ge}{\geqslant}
\renewcommand{\le}{\leqslant}

\title{Polylogarithmic equilibrium treatment of molecular aggregation and critical concentrations\footnote{Reference: Denis Michel and Philippe Ruelle. 2017. Polylogarithmic equilibrium treatment of molecular aggregation and critical concentrations. Phys. Chem. Chem. Phys. 19, 5273-5284. doi: 10.1039/c6cp08369b}}
\date{} 

\author{Denis Michel$ ^{\dagger} $ and Philippe Ruelle$ ^{\ddagger} $}

\begin{document}
\maketitle
\begin{small} $ ^{\dagger} $ Universite de Rennes1-IRSET. Campus Sant\'e de Villejean. 35000 Rennes France. Email: denis.michel@live.fr. \end{small} \\ \begin{small} $ ^{\ddagger} $ Universit\'e catholique de Louvain - Institut de Recherche en Math\'ematique et Physique. Chemin du Cyclotron, 2 B-1348 Louvain-la-Neuve, Belgium. Email: philippe.ruelle@uclouvain.be. \end{small}

\begin{multicols}{2}

\textbf{A full equilibrium treatment of molecular aggregation is presented for prototypes of 1D and 3D aggregates, with and without nucleation. By skipping complex kinetic parameters like aggregate size-dependent diffusion, the equilibrium treatment allows to predict directly time-independent quantities such as critical concentrations. The relationships between the macroscopic equilibrium constants for the different paths are first established by statistical corrections and so as to comply with the detailed balance constraints imposed by nucleation, and the composition of the mixture  resulting from homogeneous aggregation is then analyzed using the polylogarithm function. Several critical concentrations are distinguished: the residual monomer concentation in equilibrium (RMC) and the critical nucleation concentration (CNC), that is the threshold concentration of total subunits necessary for initiating aggregation. When increasing the concentration of total subunits, the RMC converges more strongly to its asymptotic value, the equilibrium constant of depolymerization, for 3D aggregates and in case of nucleation. The CNC moderately depends on the number of subunits in the nucleus, but sharply increases with the difference between the equilibrium constants of polymerization and nucleation. As the RMC and CNC can be numerically but not analytically determined, ansatz equations connecting them to thermodynamic parameters are proposed.} \\
\\
\textit{Keywords}: Nucleation; aggregation; self-assembly; noncovalent polymers; equilibrium.

\section{Introduction}

Noncovalent molecular self-aggregation is a widespread process which can have desired and pathological outcomes in biology \cite{Pham}. Nucleation-dependent aggregation is most generally investigated for its kinetics \cite{Cohen,Gillam,Michaels,Kumar}, but equilibrium approaches are appropriate to obtain time-independent properties, like detailed balance relations and critical concentrations. The notion of critical concentration is widely used in this field but not always with the same meaning. It covers different names such as the critical aggregation concentration (CAC) usually equated to the equilibrium depolymerization constant ($ K_{d} $), the critical fibril concentration (CFC), or the critical micellar concentration (CMC) whose relationships are somewhat confusing. There is a relative consensus about the critical monomer concentration, generally assumed to correspond to the concentration of free monomers in equilibrium and accordingly written here RMC for residual monomer concentration. When perturbing equilibrium by adding or removing free subunits, the aggregates grow or dissolve respectively, so in equilibrium and in presence of aggregates, the RMC corresponds to a critical concentration for aggregate growth, close to the equilibrium depolymerization constant in supersaturated mixtures \cite{Frieden,Cohen,fibrils}. Another type of critical concentration, called here CNC, describes the threshold of total subunits above which aggregation becomes significant. The CAC and CNC have been assumed identical \cite{Wegner,Howard} but this is only an approximation in case of very weak nucleation.  We propose to clarify these questions by rigorously establishing the RMC and CNC. Since concentrations evolve in a closed nonequilibrium system, critical concentrations are fixed values characteristic of the system considered and should be calculated at equilibrium. Therefore, we present a universal equilibrium model of aggregation, with and without nucleation, for 1D (fibrils) and 3D (clusters) aggregates. For simplicity, we will consider homogeneous nucleation and ignore possible additional parameters like secondary nucleation or the gravitational sedimentation of clusters. The aggregates will be supposed uniformly composed of identical elementary components $ S_{1} $, including in the nuclei. As this study considers exclusively equilibrium, aggregation will be described using equilibrium constants, which are simpler but fundamentally equivalent to energies, work and thermal dissipation \cite{Shchekin}, while avoiding exponentials and making transparent the detailed balance relationships. Kinetic equations of Smoluchowski and Becker-D\"oring will not be used to directly identify the final aggregated fraction.

\section{Specificities of equilibrium treatments}

A striking feature of nucleation-dependent aggregation is its lag phase \cite{Arosio}, which may explain why it has historically been most often studied kinetically. In kinetic approaches, the final amounts of the different molecular species are deduced as asymptotic limits at infinite time \cite{Cohen} or by cancelling all the net fluxes in Smoluchowski$ ' $s equations \cite{Shchekin}. But kinetic treatments suffer from a great complexity and the ignorance of mechanistic parameters, so equilibrium treatments may be suited to determine time-independent values, in particular in case of nucleation. Nucleation is ultrasensitive to faint variations of interfering contaminants like trace ions and impurities, but this sensitivity applies mainly to lag times and has less influence in the equilibrium state. The fundamental principle of microreversibility \cite{Lewis} in which the aggregation process is no exception, has some technical virtues in the elimination of many confounding kinetic factors. In the example of fibrils, reactions of monomer addition-withdrawal and fibril breaking-rejoining, are permanent. The effective rate constants of these reactions are very complex and different for the various molecular species present in the mixture. For instance, the separation and rebinding of broken fibrils depend on their size, their diffusion and rotational capacities, relatively to the viscosity and crowding of the medium; whereas the nature of the medium is less important for the addition-removal of monomers. Incorporating all these differences is a hard modeling task and authors had recourse to alternative techniques such as moment closure \cite{Hong}. In fact, the ultimate reactions of association-dissociation of interaction surfaces conveniently positioned to react, are the same for a monomer and a fibril apex; but the reactions allowing these appropriate positions strongly differ between the reactive species. The notion of "encounter complex" (virtual complex still unbound but spatially ready to bind) \cite{DeLisi,Shoup}, precisely corresponds to this situation. For example, restricted diffusion hinders both the formation and the disappearance of the encounter complex. Otherwise the binding rate would be simply proportional to the number of binding sites \cite{Shoup}. In case of restricted diffusion, both the forward and reverse constants include different diffusion coefficients for each type of aggregate, but these coefficients disappear from the equilibrium constants \cite{DeLisi}. Using time-independent equilibrium constants $ K $ allows to get rid of a lot of microscopic mechanisms different for each type of fibril and refractory to measurements. The parameters which seriously complicate kinetic treatments, vanish in equilibrium.

\section{Nucleation-independent aggregation}

In a closed non-isolated container of fixed volume $ V $ at room temperature, before aggregation the concentration of the monomers of unit size is the total concentration $ [S_{1}]=[S]_{0} $ and following aggregation, the mixture contains monomers and aggregates $ S_{j} $ with any number $j$ of clustered monomers,

\begin{equation} 
[S]_{0} = [S_{1}]+2[S_{2}]+3[S_{3}]+\dots = \sum_{j=1}^{V[S]_{0}}j[S_{j}].
\label{s0}
\end{equation}

This purely descriptive equation is true under all circumstances, in and out of equilibrium; but in addition in equilibrium, all these individual concentrations are fixed and mutually constrained by the generalized microreversibility, which will allow us to derive a general function for the RMC. As the relative amounts of polymers and monomers will be calculated below directly at equilibrium, in the rest of the study, all the concentrations will be understood as equilibrium concentrations. To determine the function linking the amount of polymerized to total substrate $ [S_{P}] = [S]_{0}-[S_{1}] $, it is first necessary to establish the thermodynamic relationships between the different $ S_{j} $. Two types of aggregates frequently encountered in nature are fibrils and random aggregates, which require different statistical corrections of the microscopic equilibrium constants.

\subsection{1D aggregates} \label{sec:1d}

The surfaces through which the subunits interact are generally not uniformly distributed around the subunit, but restricted to specific facets, so that the inter-molecular bonds are oriented in the aggregate. This polarized mode of interaction generates non-random structures like fibrils. The case of fibrils is simple as it does not require the statistical balancing of binding constants \cite{fibrils}. Indeed, the apex of a fibril that serves as a platform for its elongation, is independent of its size. As a consequence, the macroscopic binding constant is identical to the microscopic constant $ K $ for all the binding steps, irrespective on whether fibril elongation proceeds through one--by--one monomer addition/substraction or fibril joining/breaking. For the different elongation steps, the generalized equivalence of the back and forth fluxes gives
\begin{equation}
[S_{2}]=K[S_{1}]^{2}, \qquad [S_{3}]=K[S_{1}][S_{2}]=K^{2}[S_{1}]^3,
\end{equation}
and generally,
\begin{equation}
[S_{j}]=K^{j-1}[S_{1}]^{j}, \quad j\ge2.
\label{E:gp3}
\end{equation}
It follows from Eq.(\ref{s0}) that the total amount of polymerized substrate is given by
\begin{equation} 
[S_P] = [S]_0 - [S_1] = \frac{1}{K} \sum_{j=2}^{\infty}\; j(K[S_{1}])^{j},
\end{equation} 
where, assuming that the total available volume is very large, the upper limit of the summation has been extended to infinity. We note that this requires $K[S_1] < 1$.

\subsection{3D aggregates}

The number of noncovalent chemical bonds to form/break for associating/dissociating two monomers is embodied in the intrinsic constant and remains identical irrespective of whether these monomers are isolated or included (but accessible) in large aggregates. Hence, as for fibrils, the microscopic constants will be assumed identical for all the binding steps. The only difference is statistical. Statistical corrections of microscopic equilibrium constants are necessary when there are several binding sites per molecule. This rule which is classically used for multimeric ligand receptors, as for oxygen binding to hemoglobin, also applies to multidimensional aggregation, with the difference that the monomers are ligands when free and then become binding sites of macromolecules when clustered. Given the huge number of subunits in ordinary aggregates, we will simply assume that the binding sites available for interacting with an aggregate are proportional to its surface, because internal sites are unreachable. Conversely, a single unit can dissociate from an aggregate only if it is surface-exposed but not embedded inside the aggregate. In his pioneer equation for coagulation, Smoluchowski enumerated all association/dissociated reactions differing for the different sizes of the clusters \cite{Smoluchowski}. This exhaustive approach can be bypassed in equilibrium thanks to the principle of detailed balance, which applies to any subset of pathways in a system of any size, and according to which the back and forth fluxes are equal between any two populations of molecular species, including the rarest ones \cite{Lewis}. This rule has a practical interest in the present case since it allows to choose the simplest way (addition/subtraction of monomers) to establish the relationships between polymer concentrations. The other ways of polymerisation (fusion/fission of polymers) are automatically given by transitivity and it is not necessary to take them into account to obtain the correct results.

\subsubsection{One by one monomer addition}

The relationship between the surface and the number of components in the cluster depends on its architecture, which can be so variable (including ordered crystals or dendrimers with various branching modes), that they can not be all treated using the same statistical rule. To show an example of 3D statistical correction, we will arbitrarily select the simple theoretical case of densely packed clumps without empty cavities, owing to a perfect induced-fit mechanism. The volume of compact clusters corresponds to the number of its components of unit volume. Random particle accretion is expected to yield spherical average cluster shapes since the highest shape entropy is attained for the sphere.

In the polymerisation step $S_{j-1}+S_{1} \rightleftharpoons  S_{j}$, a monomer can either bind to an aggregate containing $j-1$ subunits, or dissociate from an aggregate containing $j$ subunits. We assume that the monomer addition or removal with or from an aggregate, is proportional to the surface of the aggregate, itself being related to its volume by an exponent 2/3.

All together, these conditions imply that the effective equilibrium binding constant is related to the intrinsic monomer-monomer binding constant $ K $ through
\begin{equation} 
K_{(j-1) \rightleftharpoons  j}= K \left (\dfrac{j-1}{j}  \right )^{2/3}.
\end{equation} 

As long as the different components of the nascent cluster are all surface-exposed (typically for $ j $ between 2 and 6), this correction is not necessary but it rapidly holds for larger complexes. In addition, small clusters can be neglected as their contribution in the total amount of aggregates is very low and further reduced in case of nucleation, as discussed later. The concentration of complexes of size $j$ follows
\begin{equation}
[S_{j}] = K \left(\dfrac{j-1}{j}\right)^{2/3}[S_{1}]\,[S_{j-1}]= K^{j-1}j^{-2/3}[S_{1}]^{j}.
\label{3d}
\end{equation}
Now the sum in Eq.(\ref{s0}) yields
\begin{equation}
[S_P] = \dfrac{1}{K}\sum_{j=2}^{\infty}\; j^{1/3}(K[S_{1}]) ^{j},
\end{equation}
with the same condition $K[S_1] < 1$.

\subsubsection{Monomer addition versus cluster fusion}

The relation between $[S_{j}]$ and $[S_{1}]$ established above for one--by--one monomer addition holds as well for all binding schemes. Let us suppose that $[S_{p+q}]$ can also form by the fusion of two smaller clusters, containing respectively $p$ and $q$ monomers, and that the equilibrium relation takes the form
\begin{equation}
[S_{p+q}] = X(p,q) \, [S_{p}] \, [S_{q}],
\label{gen}
\end{equation}
with $X(p,q)$ the corrected binding constant to be determined. The compatibility with the law (\ref{3d}) requires
\begin{equation}
X(p,q) = K \left (\frac{1}{p} + \frac{1}{q}\right )^{-2/3} = K \left(\frac{p\,q}{p+q}\right)^{2/3}.
\end{equation}

In the more general situation where the factor $j^{-2/3}$ in (\ref{3d}) is replaced by some function $F(j)$ (satisfying $F(1)=1$), the corrected constant is given by $X(p,q) = K \frac{F(p+q)}{F(p)F(q)}$. The special case $F(j)=1$ corresponds to 1D aggregates, described in Section \ref{sec:1d}.

\section{Nucleation}

Nucleation is a universal phenomenon involved in diverse areas of physics and which stimulated the developement of several theories. It is the sharp transition between different phases initially described for the condensation of droplets from saturated vapor. It results from the abrupt switch between the antagonistic tendencies of dispersion of monomers (maximizing entropy) and of stabilisation of monomers (minimizing energy). The rates of condensation, increasing with the sizes of the clusters, and that of evaporation, decreasing with the size of clusters, calculated for instance in the dynamical nucleation theory \cite{Garrett}, equalize for the critical cluster size called nucleus, which is the energetic bottleneck for generalized aggregation. These general principles can be translated in the present context, when aggregation is conditioned to the preliminary formation of small clusters, which themselves can form only above a critical concentration of the elementary units. Nucleation has both kinetic and concentration barriers, but it is most often conceived kinetically, as a stochastic event with an exponential waiting time which can be very long, unless it is shortened by catalyzing impurities. Nucleation is not really a slow process, as it is often described, but a very fast phenomenon preceded by a long waiting period. In addition to its popular lag time effect, nucleation has also an impact in the genesis of a specific critical concentration fixed by thermodynamic parameters at equilibrium. Many models of nucleation have been proposed, including critical proximity between monomers \cite{Stillinger} or energy-based clusters \cite{Harris}. Concretely, the origin of nucleation for aggregation is the insufficient number of inter-subunit bonds stabilizing the complex until a critical size is reached. If a $ n $-mer nucleus is a prerequisite for aggregate growth, one can assume that no cluster of size between 2 and $n-1$ is present at equilibrium and the total amount of fibril components spreads over the other species
\begin{equation}
[S]_{0} = [S_{1}]+n[S_{n}]+(n+1)[S_{n+1}]+ \dots +(n+j)[S_{n+j}] + \dots 
\end{equation}

The precise reactions involved in nucleation are not accessible experimentally, but the overall characteristics of this process suggest that it is a rare and sudden phenomenon, thermodynamically unfavorable compared to the subsequent aggregate growth. Nuclei are conceived here as inherently unstable but conveniently arranged for initiating polymerization. Indeed on the one hand, stable nuclei would be simply pre-assembled building blocks in the classical process of exothermic hierarchical assembly. On the other hand, oligomers made of the same components but not conveniently arranged for further polymerization can not be considered as nuclei, even if such oligomers can exist and play roles by reducing the availability of monomers \cite{Gosal,Miti}. In the simple mass action mechanism retained here, $ n $-mer nuclei will be indirectly defined by the absence of clusters containing less than $ n $ subunits, through a Hill reaction involving the direct condensation of $ n $ monomers and corresponding to the quasi-simultaneous collision between $n$ monomers, or to a chain of sequential addition of components often initiated but rarely completed, like a preferentially backward random walk with a final absorbing state locked when exothermic polymerization stabilize the fleeting nuclei \cite{fibrils}. In fact, the calculation of intermediate states and of first arrival times shows that both mechanisms are very similar. Indeed, a finite backward random walk is an endothermic stepwise process which can be occasionally completed in a probabilistic manner like a single jump \cite{stepwise}. It is therefore reasonable to select the $ n $-order binding reaction as the minimalist model of nucleation. The nucleation constant, which we will denote by $K_n$, will be defined from the basic view of nuclei, accepted in all the nucleation theories, as the particular cluster size $ n $ for which the fluxes of dispersion and accretion equalize in equilibrium \cite{Farkas}, which simply reads

\be
k_{accr}[S_1]_{eq}^n=k_{disp}[S_n]_{eq}
\ee
leading to
\be
 K_n=\dfrac{k_{accr}}{k_{disp}}=\dfrac{[S_n]_{eq}}{[S_1]_{eq}^n}
\ee

The unit of $ K_n $ is therefore that of an inverse concentration to the power $n-1$.

\subsection{Nucleation-dependent 1D aggregates}

For 1D clusters with $n$-mer nuclei, the fundamental equilibrium conditions read
\be
[S_j] = K^{j-n} [S_n] \, [S_1]^{j-n} = K_n K^{j-n} [S_1]^j, \qquad j \ge n.
\ee
The sum rule then implies 
\be
[S_P] = K_n \, K^{-n} \, \sum_{j=n}^\infty \: j \, (K[S_1])^j, 
\label{1dnuc}
\ee
and $K[S_1]<1$.

\subsection{Nucleation-dependent 3D aggregates}

The concentrations obey the following relations
\be
\begin{split}
[S_j] &= K^{j-n} \Big(\frac nj \Big)^{2/3} \, [S_n] \, [S_1]^{j-n}\\& = K_n \, K^{j-n} \, \Big(\frac nj \Big)^{2/3} \, [S_1]^j, \qquad j \ge n,
\end{split}
\ee
which lead to the following value for $[S_P]$,
\be
[S_P] = K_n \, K^{-n} \, n^{2/3} \, \sum_{j=n}^\infty \: j^{1/3} \, (K[S_1])^j,
\label{3dnuc}
\ee
with the same restriction $K[S_1]<1$. 

Comparing the expressions of $[S_P]$ with and without nucleation, one sees that the nucleation manifests itself in two ways: by the value of $n$ and by the way the nucleation constant $K_n$ compares with $K^{n-1}$ (modulated by a factor $n^{-2/3}$ for 3D aggregates). This suggests to trade the nucleation constant $K_n$ for the more convenient dimensionless parameters,
\be
\begin{split}
& F_n \equiv K_n \, K^{1-n} \quad \text{(1D models)}, \\ & G_n \equiv n^{2/3} \, K_n \, K^{1-n} \quad \text{(3D models)}.
\end{split}
\ee
We will assume that these two parameters take their values in $[0,1]$, considering that classical nucleation cannot facilitate aggregation.

The strict absence of nucleation is obtained by setting $n=2$ as well as $F_2=1$ or $G_2=1$. The effects of nucleation can be enhanced by choosing larger values of $n$ and/or by taking smaller values of $F_n,G_n$. Changing the value of $n$ while keeping $F_n$ or $G_n$ close to 1 merely changes the lower bound of the summation over $j$, meaning that clusters of small (and bounded) sizes are absent. As the qualitative behaviour of $[S_P]$ is expected to be dominated by the vast majority of larger clusters, the precise value of $n$ should not be very important. It suggests that the parameters $F_n,G_n$ are what really determines the amount of nucleation. 

The following analyses will be made for general values of $n$ and $F_n,G_n$, allowing for a continuous fine-tuning of nucleation. Indeed, the total absence of nucleation is very unlikely in the homogeneous aggregation, and there is actually a continuum between minimal nucleation ($F_n,G_n \sim 1$) and extreme nucleation ($F_n,G_n \sim 0$). 

\subsection{Relationships between the constants imposed by nucleation under the detailed balance}

Equilibrium modeling must comply with the law of generalized micro-reversibility which implies that the concentration of every molecular species should be equivalently calculated regardless of the channel used to build this species. This rule, that is an extension of the Wegscheider's condition \cite{Wegscheider} and the generalized microreversibility \cite{Lewis}, allows to establish the relationships linking together the different constants of the network. For 1D aggregates, the $ [S_j] $ complex can be obtained either (i) by successive additions of ($ j-n $) monomers on a nucleus of $ n $ subunits,
\begin{subequations}\label{E:gp}
\begin{equation} 
[S_j]=K \, [S_1] \, [S_{j-1}],
\end{equation} \label{E:gp1}
or (ii) by combining two existing complexes of $ i $ and $ (j-i) $ subunits ($ n <i <j $). 
\begin{equation} 
[S_j]=K \, [S_i] \, [S_{j-i}] \, K^{n-1}/K_n.
\end{equation} \label{E:gp2}
\label{wr}
\end{subequations}
As shown in Eq.(\ref{wr}), the equivalence between these pathways implies that the microscopic constant of polymer joining must be corrected by a factor $ K^{n-1}/K_n $. The physical meaning of this correction is clear: to take into account a single nucleation per polymer, a nucleation is substituted by the polymerization of an equivalent number of subunits. For 3D aggregates in addition to this rule, the statistical correction imposed by the number of accessible binding sites still applies, giving for any $ p $ and $ q $ larger than $ n $,
\begin{equation} 
[S_{p+q}]=\dfrac{K^{n}}{K_n}\left(\frac{pq}{p+q}\right)^{2/3} [S_p] \, [S_q].
\end{equation} 

\section{The polymerized versus total substrate relationships}

Our main interest is in the amount $[S_P] = [S]_{0} - [S_1]$ of aggregated substrate. It is natural and convenient to rewrite the main equations in terms of dimensionless variables. So in addition to using the parameters $F_n$ and $G_n$, we will work with the following dimensionless variables,
\be
x \equiv K[S]_{0}, \qquad y \equiv K[S_P], \qquad x-y \equiv K[S_1] < 1.
\ee

The relations (\ref{1dnuc}) and (\ref{3dnuc}) are then equivalent to the following non-linear equations,
\begin{subequations}
\begin{equation} 
y = F_n \times \Big\{ \li_{-1}(x-y) - \sum_{j=1}^{n-1} \: j\,(x-y)^j \Big\} 
\label{y1d}
\end{equation} 
for 1D models,
\begin{equation} 
y = G_n \times \Big\{ \li_{-1/3}(x-y) - \sum_{j=1}^{n-1} \: j^{1/3}\,(x-y)^j \Big\}, 
\label{y3d}
\end{equation} 
\end{subequations}
for 3D models. In both cases, we have used the polylogarithmic functions $\li_s(z)$, defined by
\be
\li_s(z) = \sum_{j=1}^\infty \: \frac{z^j}{j^s}, \qquad |z|<1.
\label{poly}
\ee

As our main purpose is to compute $[S_P]$ in terms of $[{S}]_0$, the mathematical problem is to extract the value of $y=y(x)$ as a function of $x$ from the above relations. From their form and the functions involved, one expects this to be a highly non-trivial if not hopeless task for general $n$. Even in the simpler 1D case, for which the function $\li_{-1}$ is elementary (see below), Eq.(\ref{y1d}) can be seen to be equivalent to a polynomial equation in $x-y$ of degree $n+1$ if $F_n<1$, and of degree $n$ if $F_n=1$. So the only really simple case is the 1D case without nucleation ($n=2$ and $F_2=1$), which reduces to a quadratic equation. We will briefly discuss it explicitly below, as an illustration. All 3D cases are expected to be hard as the function $\li_{-1/3}$ is not elementary at all. In all the hard cases, only perturbative solutions in the form of series may be obtained.

Before presenting some details of our analysis, we briefly recall some basic features of polylogarithmic functions \cite{nist}. 

\subsection{Polylogarithms} \label{polylogs}

Polylogarithms are functions of a complex variable $z$ depending on a complex parameter $s$. In the special case $s=1$, it reduces to the usual logarithm, $\li_1(z) = -\log{(1-z)}$. For any complex $s$, the series (\ref{poly}) is convergent for $|z|<1$ and defines an analytic function inside the unit circle; it also converges on the unit circle $|z|=1$, provided ${\rm Re} \, s >1$. Outside the unit circle, $\li_s(z)$ is defined by analytic continuation, generally takes complex values and may be multi-valued (existence of branch cuts). 

The functions associated to values of $s$ differing by integer gaps are differentially related, on account of
\be
\li_{s-1}(z) = z\frac{\rm d}{{\rm d} z} \li_s(z).
\ee

By successively applying this relation to $\li_1(z)$, one easily sees that $\li_s(z)$ is an elementary rational function of $z$ when $s$ is a negative integer, with a single pole on the entire complex plane, of order $1-s$ at $z=1$,
\be
\begin{split}
& \li_{0}(z) = \frac{z}{1-z},\\ & \li_{-1}(z) = \frac{z}{(1-z)^2},\\
& \li_{-2}(z) = \frac{z(1+z)}{(1-z)^3}.
\end{split}
\ee

More generally for $s<1$ real, $\li_s(z)$ is singular at $z=1$.  By using the discrete Tauberian theorem \cite{Feller}, the main divergence can be computed explicitly when one approaches the singularity from below on the real axis ($\Gamma(s)$ is the Euler Gamma function),
\be
\li_s(z) \sim \frac{\Gamma(1-s)}{(1-z)^{1-s}}, \qquad z \to 1^-.
\label{main}
\ee
It also has lower order singular terms, as shown by the following expansion, valid in the left real neighbourhood of $1$ and for $s<1$ real and non-integer,
\be
\li_s(z) = \frac{\Gamma(1-s)}{(1-z)^{1-s}} \cdot A_s(1-z) + B_s(1-z), \qquad z \to 1^-,
\label{exp}
\ee
where $A_s(z)$ and $B_s(z)$ are regular Taylor series ($\zeta(s)$ is the Riemann zeta function)
\begin{subequations}
\begin{equation}
\begin{split}
& \hspace{-1cm} A_s(x) = 1 + {\textstyle \frac 12}(s-1) x \\ & + {\textstyle \frac 1{24}}(s-1)(3s+2) x^2 + \ldots 
\end{split}
\end{equation}
\vspace{-5mm}
\begin{equation}
\begin{split}
& \hspace{-1cm} B_s(x) = \ \zeta(s) - \zeta(s-1) \, x \\& + {\textstyle \frac 12} [\zeta(s-2) - \zeta(s-1)] \, x^2 + \ldots
\end{split}
\end{equation}
\end{subequations}

\subsection{Asymptotic solutions}

The basic properties of polylogarithms recalled in the previous section allow us to characterize the solutions to both Eqs (\ref{y1d}) and (\ref{y3d}) in two asymptotic regimes: when the initial concentration is large ($x \gg 1$), and when the renormalized nucleation constants $F_n,G_n$ tend to 0. 

The variables $x$ and $y$ take positive real values, but they are not independent since their difference $x-y$ is bounded by 1. So if we let $x$ become unboundedly large, so must be $y$ and therefore also the right-hand sides of Eqs (\ref{y1d}) and (\ref{y3d}). Because the finite sums over $j$ remain bounded, it follows  that the functions $\li_{-1}(x-y)$ and $\li_{-1/3}(x-y)$ must diverge, and this means that their argument $x-y$ comes close to 1. Therefore when $x$ is large, $x-y$ is asymptotically equal to 1, that is, the solutions $y(x)$ are asymptotically linear, in both models, 1D or 3D,
\be
y(x) = x - 1 + o(1), \qquad \text{for $x$ large},
\label{rmc}
\ee
where the correction $o(1)$ is a function of $x$ that goes to 0 when $x$ goes to infinity. As shown later in Section 6, it is typically given by a series of fractional powers of $x^{-1}$, which can be computed perturbatively to any finite order. It is however model-dependent. The dominant correction is proportional to $x^{-1/2}$ for the 1D models, to $x^{-3/4}$ for the 3D models.

Let us now look at the solutions of Eqs (\ref{y1d}) and (\ref{y3d}) when $F_n$ and $G_n$ tend to 0, that is, in the case of extreme nucleation. In that limit, and if we keep $n$ fixed, the finite sums over $j$ can be neglected since they are bounded by a constant only depending on $n$, which is multiplied by $F_n$ or $G_n$. So the resulting equations reduce to $y = F_n \, \li_{-1}(x-y)$ and $y = G_n \, \li_{-1/3}(x-y)$. In the process of taking the limit, one should bear in mind that the solution $y=y(x)$ depends implicitly on $F_n,G_n$. Let us denote its limit by $y_0(x)$, namely $y_0(x) = \lim_{F_n,G_n \to 0} y(x)$.

Let $x$ be such that $y_0(x)>0$. It implies that the limit of $\li_{-\alpha}(x-y)$, with $\alpha=1$ or $1/3$, diverges (at the proper rate), namely that the limit of its argument is equal to 1. As above this readily yields $y_0(x) = x-1$. However for $x<1$, this solution contradicts our positivity assumption $y_0(x) > 0$. Therefore we obtain that the solution $y_0(x)$ in the limit of extreme nucleation is piece-wise linear, with a discontinuous change of slope at $x=1$, again for both 1D and 3D models,
\be
y_0(x) = \begin{cases}
0 & \text{for } x \le 1,\\
x-1 & \text{for } x \ge 1.
\end{cases}
\label{extreme}
\ee

\subsection {1D aggregation without nucleation}

Setting $n=2$ and $F_2=1$ in Eq.(\ref{y1d}) and using the explicit form of $\li_{-1}(z)$ yields the following simple equation,
\be
x = \li_{-1}(x-y) = \frac{x-y}{(1-x+y)^2},
\ee
equivalent to a quadratic equation for $y$. The solution which is positive for every positive $x$ is given by
\be
y(x) = x - 1 - \frac{1}{2x} + \sqrt{\frac1x \left(1 + \frac1{4x}\right)}.
\ee
Its asymptotic form, for large $x$, is indeed given by the linear function $x-1$, with a first dominant correction proportional to $1/\sqrt{x}$, as announced in Section 5.2. The function $y(x)$ is regular, $y(0)=0$, and smoothly converges to its linear asymptotic form as $x$ increases.

\section{Perturbative solutions for the RMC and CNC}

As noted earlier, the residual monomer concentration $[S_1]$ cannot be determined as an explicit function of the initial monomer concentration $[S]_0$ and the aggregation parameters. However for large $x$, a solution can be written in the form of a infinite series in inverse powers of $x$, which is computable to any desired order. As to the critical nucleation concentration, informally described as the monomer concentration threshold beyond which aggregation is important, it has so far not been given a precise mathematical definition. In the following, we make a definite proposal for such a definition, and determine its value as a function of the nucleation parameters, again, in the form of aymptotic series.

\subsection{The RMC} \label{RMC}

The residual monomer concentration $[S_1] = K^{-1}(x-y)$ is asymptotically equal to $K^{-1}$, namely in the limit of an infinite total concentration of subunits $[S]_0=K^{-1}x$, see Eq.(\ref{rmc}). As the concentration of free subunits can be measured either directly or by subtraction using the plots showing aggregated versus total concentrations, it is of interest to evaluate the difference between the RMC and $1/K$. Since the difference precisely vanishes in the limit of large $x$, we make the Ansatz that the corrections are given by inverse powers of $x$, which may be computed iteratively in a rather simple way.

So we assume the first correction to be of the form $y = x - 1 + \frac a{x^\alpha} + \ldots$, with $\alpha>0$. We plug this form in the identity we want to solve, either Eq.(\ref{y1d}) or Eq.(\ref{y3d}), and determine the values of $a$ and $\alpha$ in order to satisfy the identity at the dominant order. For the 3D models for instance, we obtain, from (\ref{main}),
\begin{equation}
\begin{split}
y_{3{\rm D}} & = x - 1 + \frac a{x^\alpha} + \ldots \\ & = G_n \Big\{ \li_{-1/3}\big(1 - \frac a{x^\alpha} + \ldots\big)\\& \qquad\qquad\qquad  - \sum_{j=1}^{n-1} \: j^{1/3}\,  \big(1-\frac a{x^\alpha} + \ldots\big)^j \Big\} \\ &
= G_n \Big\{ \Gamma({\textstyle\frac43}) \, \Big(\frac{x^{\alpha}}{a}\Big)^{4/3} + \ldots + \sum_{j=1}^{n-1} \: j^{1/3} + \ldots \Big\}.
\end{split}
\end{equation}
The dominant term on either side should match, implying $x = G_n \Gamma({\textstyle\frac43}) \, \big(\frac{x^{\alpha}}{a}\big)^{4/3}$. Hence $\alpha=\frac34$ and $a = [G_n \Gamma(\frac43)]^{3/4}$.

The next correction, assumed to have the same form (but with different exponent and coefficient), is determined in the same way. In fact, further correction terms, as many as we want, can be similarly computed, one by one, for the 1D and 3D models. These calculations only require to know the series expansions of the functions $\li_{-1}(z)$ and $\li_{-1/3}(z)$ for $z$ close to $1^-$, as given in (\ref{exp}), and usually integrated in symbolic mathematical computation programs. The first few corrections are given by 
\begin{subequations}
\begin{equation}
\begin{split}
y_{1{\rm D}}= & \quad x - 1 + \frac{\sqrt{F_n}}{x^{1/2}} \\ & - \frac{F_n}{2x} - \frac{\sqrt{F_n} \big[F_n(2n^2-2n-1)-4\big]}{8x^{3/2}} + \ldots
\end{split}
\end{equation}
\begin{equation}
\begin{split}
y_{3\rm D} = & \quad x - 1 + \frac{\big[G_n \Gamma(\frac43)\big]^{3/4}}{x^{3/4}} - \frac{\big[G_n \Gamma(\frac43)\big]^{3/2}}{2x^{6/4}}\\& + \frac{3\big[G_n \Gamma(\frac43)\big]^{3/4}\big[G_n\zeta(-\frac13) -G_n H_{n-1}^{(-\frac13)}-1\big]}{4x^{7/4}} + \ldots
\end{split}
\end{equation}
\end{subequations}
where $H_n^{(\alpha)} = \sum_{j=1}^n j^{-\alpha}$ are generalized harmonic numbers.

Although these expansions are not particularly illuminating, one can nevertheless make a few instructive observations. Because of the different exponents characterizing the first correction, $3/4$ against $1/2$, the convergence to the asymptotic value $y = x-1$ is a bit faster for the 3D models than for the 1D models. In both models, the convergence is also faster for stronger nucleation (i.e. for smaller nucleation constants $F_n$ or $G_n$), the effect being again strengthened for the 3D models. Finally, we note that the explicit dependence in $n$ (namely, apart from the nucleation constants themselves) is rather weak, since it only shows up in the third correction. These rather strong convergence rates validate the accepted equivalence between the RMC, the CAC and $1/K$ in all cases of aggregation in supersaturated solutions.

\subsection{The CNC}

When nucleation effects are strong enough, one observes that the concentration of aggregates $[S_P]$, as a function of the monomer concentration $[S]_0$, remains close to zero up to a certain value of $[S]_0$, after which $[S_P]$ raises significantly. This is illustrated in a typical example by the blue curve in Fig.1. The threshold value of $x$, which we will denote by $x^*$, is in this example close to $0.8$. The corresponding concentration, given by $x^*/K$, is precisely the critical nucleation concentration (CNC).
Looking at the plots, one can see that the concentration (in blue) shows a sort of smoothed-out angle (it gets sharper as the nucleation increases), where the curvature is higher than in the other portions of the plot. One also sees that the maximum of the curvature (in grey) adequately locates the threshold.

\begin{center}
\includegraphics[width=8.5cm]{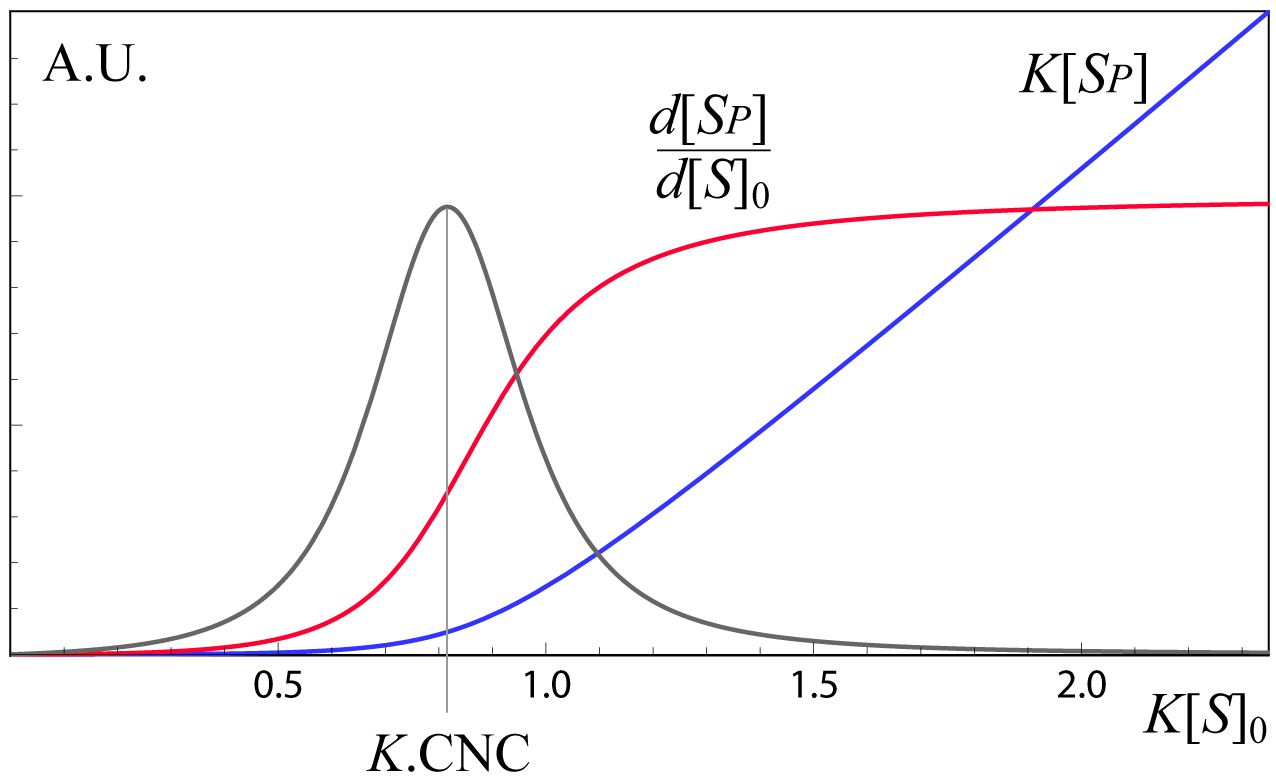} \\
\end{center}
\begin{small} \textbf{Figure 1} Representation in arbitrary units of the accumulation of aggregates (blue curve), its derivative (red curve) and its curvature (grey curve) as functions of the total concentration of subunits. The nucleation parameters used are $n=3$ and $F_3=0.004$. The curvature is used to determine the critical nucleation concentration, as defined in (\ref{curv}), over which aggregates significantly accumulate. $ K $ is the binding constant of polymerization. \end{small}\\

This strongly suggests to define the threshold $x^*$ and the corresponding CNC as the point where the concentration curve  has a maximal curvature. Mathematically the curvature of a function at a point $x$ is defined as the inverse radius of the osculating circle at $x$, namely the circle that best fits the curve locally \cite{curv}. The curvature $\gamma(x)$ of a function $y(x)$ can be computed from the following expression,
\be
\gamma(x) = \frac{y''(x)}{[1 + y'^2(x)]^{3/2}},
\label{curv}
\ee
and is positive where the function is convex, negative where it is concave. In the present case, the concentration function is convex everywhere, so that its curvature is positive. 

We are therefore led to propose the following definition of the Critical Nucleation Concentration,
\be
\text{CNC =$\frac{x^*}{K}$, where $x^*$ satisfies :} \quad \gamma'(x^*) = 0.
\ee
Evidently, we reject the solutions $x^*=0$ and $x^*=+\infty$, where the curvature is minimal (and equal to 0).

As one does not have a explicit expression of the function $y(x)$, it seems very unlikely that we will be able to find one for $x^*$. But since the threshold $x^*$ is significantly different from 0 only when the nucleation effects are noticeable, one may try to find $x^*$ as a series in positive powers of the nucleation constants $F_n,G_n$, assumed to be small. In the limit of extreme nucleation $F_n,G_n \to 0$, we know from (\ref{extreme}) that the threshold is equal to $x^*=1$ (the curvature is infinite), in both families of models. Therefore we look for series expansions starting off like $x_{\rm 1D}^* = 1 - a_1 F_n^{\alpha_1} + \dots$ in the 1D models, and $x_{\rm 3D}^* = 1 - b_1 G_n^{\beta_1} + \dots$ in the 3D models, with $a_1,\alpha_1,b_1,\beta_1$ positive real numbers. Some details are given in the Appendix as to how these can actually be computed, along with several higher order terms. We only give here the final results. 

In the 1D models, we obtain the following expansion in powers of $F_n^{1/3}$, 
\begin{subequations}
\label{xstar1d}
\be
\begin{split}
x^*_{\rm 1D}& = 1 - a_1 \,F_n^{1/3} - a_2 \, F_n^{2/3} - a_3 \, F_n\\
& - \frac{n(n-1)}{2} \, F_n \exp\Big\{\!-\!\frac{(2n-1)}{3} \, c_1 \, F_n^{1/3}\Big\} + {\cal O}(F_n^{4/3}),
\end{split}
\ee
with the coefficients given by
\bea
a_1 &\simeq& 1.0666, \qquad a_2 \simeq 0.2177,\\
\noalign{\smallskip}
a_3 &\simeq&  0.0948,\qquad c_1 \simeq 1.507.
\eea
\end{subequations}

More terms can be easily computed if needed, but their explicit exact values (as those given in the Appendix) are increasingly complicated. For the values of the parameters used in Fig.1, namely $n=3$ and $F_3=0.004$, the series truncated as above to the first three non-trivial terms yields $x^*_{\rm 1D} \simeq 0.812824$.

For the 3D models, the coefficients in the expansions are somewhat more complicated, reflecting the higher complexity of the $\li_{-1/3}$ function. We find
\begin{subequations}
\label{xstar3d}
\be
\begin{split}
x^*_{\rm 3D} &= 1 - b_1 \,G_n^{3/7} - b_2 \, G_n^{6/7} - b_3 \, G_n\\
& - H_{n-1}^{(-\frac13)} \, G_n \exp{\Big\{\!-\!\frac{H^{(-4/3)}_{n-1}}{H^{(-1/3)}_{n-1}}\, d_1\, G_n^{3/7}\Big\}} + {\cal O}(G_n^{9/7}),
\end{split}
\ee
with
\bea
b_1 &\simeq& 0.6617, \qquad b_2 \simeq 0.4314,\\
\noalign{\smallskip}
b_3 &\simeq& 0.2773, \qquad d_1 \simeq 1.9509,\\
&& \hspace{-8.5mm} H_{n-1}^{(-\frac13)} = \sum_{j=1}^{n-1} \, j^{1/3}, \qquad H_{n-1}^{(-\frac43)} = \sum_{j=1}^{n-1} \, j^{4/3}.\phantom{12345}
\eea
\end{subequations}

These results could help deducing the most elusive constant of the process, $K_{n}$, from the $[S_P]$ versus $[S]_0$ plot, provided $[S]_0^{*}$, $K$ and $n$ are obtained through alternative means. To do this, an important issue is to be able to estimate the expected domain of validity of the truncated series (\ref{xstar1d}) and (\ref{xstar3d}) with respect to the aggregation parameters. 

The series involve increasing powers of the parameter $F_n$ or $G_n$, assumed to be small. In the favourable cases, namely when the coefficients decrease sufficiently fast, such series are expected to have a fast convergence. This means that the series truncated to their first three non-trivial terms, like what we have here, should yield accurate results for values of $F_n,G_n$ in a typical range $[0,0.1]$. For $F_n,G_n \sim 0.1$, the third terms $a_3F_n$ and $b_3G_n$ contribute for a few hundredths to the value of $x^*$ (which is between 0 and 1), so that the next order, respectively proportional to $F^{4/3}$ and $G_n^{9/7}$, would contribute for a fraction of that. An accuracy of a few percents could therefore be expected.

However this line of reasoning is overlooking the terms which are explicitly depending on $n$. Indeed in addition to the linear terms we have just discussed, each series contains a second linear term, explicitly depending on $n$, whose coefficients are much larger, especially in the 1D case: $n(n-1)/2$ is 30 times larger than $a_3$ for $n=3$ and 100 times larger for $n=5$ ! The corresponding ratios are respectively close to 8 and 20 in the 3D case. Even though the exponential factor corrects this, the contribution coming from the $n$-dependent term remains important.

The best way to evaluate the accuracy of the series is to compare them with the values of $x^*$ obtained numerically, that is, by solving the maximal curvature condition numerically. Although numerical, they can be considered as being exact results. A graphical illustration of such a comparison is given in Fig.2 for the 1D models.

\begin{center}
\includegraphics[width=8.5cm]{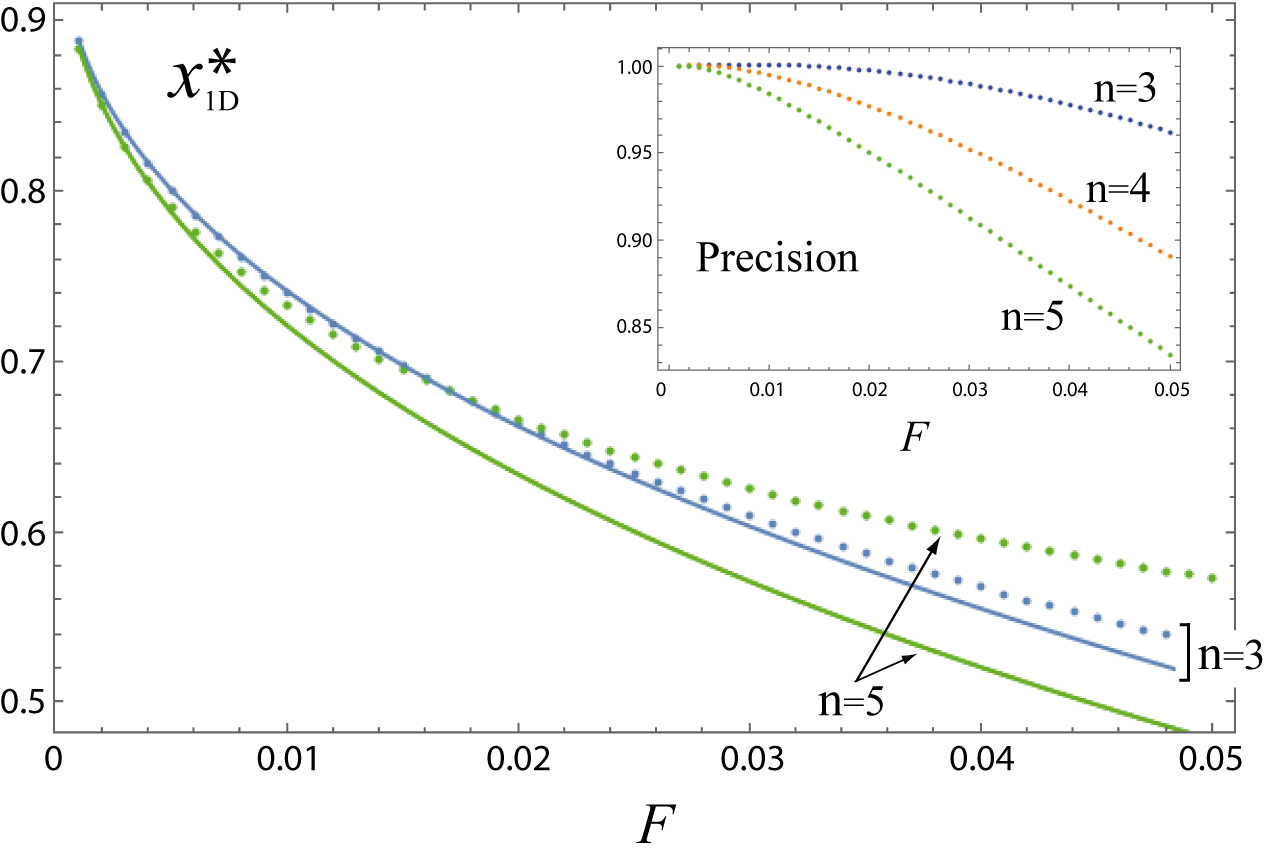}  \\
\end{center}
\begin{small} \textbf{Figure 2} Comparison, for the 1D models, between the exact numerical results (dotted) and those obtained from the series (\ref{xstar1d}) (continuous line) for the point $x_{1\rm D}^*$ of maximal curvature, defining the CNC. The blue and green curves are for $n=3$ and $n=5$ respectively. The inset shows the level of precision as a function of $F$ (the data shown are the absolute values of the ratios of $x_{\rm 1D}^*$ computed from the truncated series to its exact numerical value). \end{small}\\

From the analysis of these data, one may draw a number of conclusions. First, one cannot drop the $n$-dependent term altogether, as this would affect the accuracy for very small values of the couplings $F_n,G_n$ where the series should perform best. Second, the exponential factors are absolutely crucial to keep the deviations under control. Finally, if we want to keep an error level below 5\% in the 1D models, the values of the parameter $F_n$ should be smaller than 0.06 for $n=3$, 0.03 for $n=4$ and 0.02 for $n=5$.

In the 3D models, and because the coefficient of $n$-dependent term is smaller (the dependence on $n$ is weaker), the truncated series in (\ref{xstar3d}) yields better results than in the 1D case. For the same accuracy level (at least 95\%), the maximal allowed values for $G_n$ are 0.1 for $n=3$, 0.06 for $n=4$ and 0.045 for $n=5$. 

\section{Supplemental refinements}

The values of RMC and CNC determined above include equilibrium constants and therefore are naturally sensitive to changes of physico-chemical conditions such as temperature. In addition, compounds interfering with these processes can be added to the mixture. Dissolving agents like disaggregative chaperones are expected to shift the CNC to larger substrate concentrations and conversely stabilizing agents are expected to shift the CNC to smaller substrate concentrations. An other feature likely to be general in this field is the succession of consecutive nucleation-growth processes. 

\subsection{Secondary nucleation and nested agregation}

Secondary nucleation has kinetic \cite{Cohen2,Michaels2}, but also equilibrium signatures. The basic substrate can first generate nucleation-dependent primary polymers which can themselves become, over a new CNC, the building blocks for higher order assembly. Such nested aggregation would give rise to several angles in the $[S_P]$ vs $[S]_0$ plot, but which may be unnoticed for technical reasons, such as the detection sensitivity, for example because of inefficient binding of thioflavin T (ThT) on small polymers or too large scale separation between the primary and the secondary polymers. In case of successive nucleation-growth steps, the definition of the subunits would evolve in the successive stages, so that it is recommended to study them one by one. For the first stage, the equilibrium constants used in the  RMC$_{1} $ and  CNC$_{1} $ are defined with respect to the concentrations and binding constants of the elementary substrate $ S $, while for the next process, the RMC$_{2} $ and CNC$_{2} $ should be defined with respect to the concentration and binding constants for the primary polymers, etc.

\subsection{Kinetic vs equilibrium point of views}

The equilibrium approach is a convenient shortcut to determine the critical concentrations and is experimentally transposable to in vitro conditions. The aggregation of all the supersaturated proteins is inevitable, sooner or later, regardless of the difficulty of nucleation. Hence, considering that aggregation-prone proteins are numerous in ordinary conditions \cite{Ciryam}, in particular in the brain subject to this type of pathology, the observed aggregation diseases appear in fact relatively rare. Fortunately our brain is not in equilibrium but is an open system with permanent molecule turnovers and containing many additional molecules interfering with aggregation phenomena, like chaperones. In addition, at our lifetime scale, these phenomena are governed more by kinetic than by thermodynamic laws, that is to say a matter of rate constants and of activation energies rather than of equilibrium constants. For example although this is predicted by energy differences ($ \Delta G $), this paper does not ignite spontaneously and the building in which we stand does not immediately collapse, thanks to kinetic restrictions. Nucleation is better known for its lag time effect resembling a waiting time before a stochastic event \cite{Heneghan,Haymet}. Hence, the critical nucleation concentration would have no meaning kinetically since this stochastic event is characterized by an exponential waiting time for all substrate concentrations. If writing the mean waiting time $ \left \langle T \right \rangle $, the probability that nucleation occurs increases with time $ t $ according to $ P(\textup{nucleation before} \ t)=1-\large{\textup{e}}^{-t/\left \langle T \right \rangle} $. The models of nucleation described previously ($ n $-order reaction or backward random walk \cite{fibrils}) are particularly suited to explain a very large $ \left \langle T \right \rangle $ and that the probability of nucleation long remains close to zero. At the very microscopic level, the most realistic mode of nucleation is a random walk with many micro-steps (say $ h $) including single noncovalent bond additions. We showed that the general formula for the time of first arrival to the end of a chain with $ h $ consecutive events is \cite{steady-state} 
\begin{equation} \left \langle T \right \rangle=\sum_{i=0}^{h-1}\sum_{j=0}^{h-i-1}\frac{1}{d_{j}}\prod_{k=j}^{i+j}\frac{d_{k}}{u_{k}} 
\end{equation}
where in the present context the $ u $ are the upstream rates of the $ h $ micro-reactions necessary to build the nucleus (of which $ n $ are pseudo-first order constants including the concentration of the free subunits) and the $ d $ are the corresponding dowstream rates. This value of $ \left \langle T \right \rangle $ dramatically increases when the $ d $ slightly rise above the $ u $ \cite{stepwise}. This principle applies well to the case of nuclei whose dismantlement (as long as they are not complete) is much more probable than their completion, in a ratio corresponding to the product of the consecutive $ d $ over the product of the consecutive $ u $ \cite{stepwise}. 

\section{Conclusion}

The treatments introduced here provide a general framework for studying the oucomes of molecular aggregation, applying as well to orderly built noncovalent crystals with precise geometric meshes and to isotropic agglutination through weak nonspecific bonds.  Introducing polylogarithm and Riemann's zeta functions is uncommon in chemistry and adds new tools to the variety of approaches already proposed for modeling aggregation. We suggest that they are particularly appropriate for modeling equilibrium aggregates. Note that intriguingly, these functions have been studied in an old article entitled: "On a function which occurs in the theory of the structure of polymers" \cite{Truesdell}, but polymers were strangely absent from this paper. The present study is likely to recover the missing link between the title and the content of this puzzling article.\\
Equilibrium approaches can sometimes be somewhat theoretical for in vivo time scales, but they are nevertheless the appropriate tools to determine thermodynamic quantities such as critical concentrations. $ 1/K=K_{d} $ is a roughly acceptable approximation of the traditional critical aggregation concentration (CAC) when starting from a solution saturated enough. Accordingly, a definition of supersaturation could be $ [S]_{0}>1/K $. More precise appraisals are presented here with respect the first onset of aggregates, the behaviour of the monomer concentration  and the impact of nucleation. In addition, as the RMC and CNC can not be explicitly formulated, we offer asymptotic formulas very close to their real values. \\

\begin{center}
\huge{Appendix}
\end{center}

\appendix
\setcounter{equation}{0}  
\numberwithin{equation}{section}
\section{Maximal curvature equation}

In this Appendix, we discuss the condition of maximal curvature associated to our definition of the CNC and indicate how the asymptotic solutions (\ref{xstar1d}) and (\ref{xstar3d}) can be obtained. 

The functions of which we want to compute the curvature are solutions to the non-linear equation (\ref{y1d}) and (\ref{y3d}). Their right-hand sides involve the polylogarithmic functions truncated from below, which we will henceforth denote by $\tli_{s}(z)$ to indicate the truncation,
\be
\tli_{s}(z) \equiv \sum_{j=n}^\infty \frac{z^j}{j^s} = \li_s(z) - \sum_{j=1}^{n-1} \frac{z^j}{j^s}.
\label{tli}
\ee
To keep the notation as light as possible, the level of truncation, $n$, is assumed to be fixed throughout and not explicitly displayed. The truncated polylogarithms satisfy the same differential relation as the usual polylogarithms,
\be
\tli_{s-1}(z) = z\frac{\rm d}{{\rm d} z} \tli_s(z).
\label{diff}
\ee
The two defining equations take the form
\be
y = A \: \tli_{-\alpha}(x-y),
\label{li}
\ee
with $\alpha=1$ (with $A \equiv F_n$) or $\frac13$ (with $A \equiv G_n$).

The curvature at $x$ of the function $y=y(x)$ is given by
\be
\gamma(x) = \frac{y''(x)}{[1 + y'^2(x)]^{3/2}},
\ee
where $y$, as a function of $x$, is given by (\ref{li}).
The curvature is maximal at the point $x^* > 0$ satisfying
\be
[1 + y'^2] \, y''' - 3\,y'y''^2 = 0.
\label{max}
\ee

As a first step, it is convenient to trade the function $y(x)$ for $t(x) = x - y(x)$, which then satisfies 
\be
x - t = A \, \tli_{-\alpha}(t).
\label{maxt}
\ee 
One also has $y'=1-t', \, y''=-t''$ and $y'''=-t'''$. By using the previous equation as well as the differential relations (\ref{diff}), all derivatives of $t(x)$ can be expressed in terms of $t$ only. One finds
\bea
t' &=& \frac{t}{t + A\,\tli_{-\alpha-1}(t)},\\
\noalign{\smallskip}
t'' &=& -A t \,\frac{\tli_{-\alpha-2}(t) - \tli_{-\alpha-1}(t)}{[t+A\,\tli_{-\alpha-1}(t)]^3},\\
\noalign{\smallskip}
t''' &=& -A t \, \frac{\tli_{-\alpha-3}(t) - 3 \tli_{-\alpha-2}(t) + 2 \tli_{-\alpha-1}(t)}{[t+A\,\tli_{-\alpha-1}(t)]^4} \nonumber\\
&& \hspace{5mm} + \; 3A^2 t \,\frac{[\tli_{-\alpha-2}(t) - \tli_{-\alpha-1}(t)]^2}{[t+A\,\tli_{-\alpha-1}(t)]^5}.
\eea
Plugging these expressions into the extremum condition (\ref{max}) leads to the following equation,
\be
\begin{split}
\big[t^* + A\, &\tli_{-\alpha-1}\big] \big[\tli_{-\alpha-3} - 3 \tli_{-\alpha-2} + 2 \tli_{-\alpha-1}\big]   \\
& \times \Big[t^{*2} + 2At^* \,\tli_{-\alpha-1} + 2 \big(A \,\tli_{-\alpha-1}\big)^2\Big]\\
= 3A \,& \big[\tli_{-\alpha-2} - \tli_{-\alpha-1}\big]^2\\
& \times \Big[t^{*2} + 3At^* \,\tli_{-\alpha-1} + 2 \big(A \,\tli_{-\alpha-1}\big)^2\Big],
\end{split}
\label{tstar}
\ee
where the argument of all $\tli$ functions is $t^*$. For fixed $A$ and fixed $n$, the unique solution $0 < t^* < 1$ of this equation yields the point $x^*$ by the relation (\ref{maxt}), namely $x^* = t^* + A \, \tli_{-\alpha}(t^*)$.

The previous equation is certainly too complicated to be solved analytically for generic values of $A$, but one can hope to be able to solve it perturbatively when $A$ takes extreme values. The most interesting case is when $A$ is small, namely in the strong nucleation regime. In the limit $A \to 0$, we have from (\ref{extreme}) that $t^* = 1$ ($x^*=1, y(x^*)=0$). We therefore look for a solution $t^* = 1 - \ldots $ as a series in positive powers of $A$. The terms in the series are computed iteratively by a method similar to the one we used in Section \ref{RMC}. We illustrate it in the 1D case, $\alpha=1$.

We note that the equation (\ref{tstar}) is cubic in $A$, whose four coefficients are functions of $t^*$. Each coefficient has a series expansions in $1-t^*$, see Section \ref{polylogs}. Setting $t^* = 1-u$ and keeping the dominant term in each of the four coefficients yields
\be
\frac{24}{u^5} + \frac{36}{u^8} A - \frac{264}{u^{11}} A^2 - \frac{480}{u^{14}} A^3 = 0.
\ee
If we think of $u$ as an infinite series in $A$, namely $u = c_1 A^\gamma + \ldots$, we find that $\gamma = \frac13$ and that the coefficient $c$ must a root of
\be
(2c_1^6 - c_1^3 - 20)(c_1^3 + 2) = 0,
\ee
which we require to be real and positive (to ensure $t^* < 1$). The previous polynomial equation has only one such root, given by 
\be
c_1 = \Big(\frac{1+\sqrt{161}}{4}\Big)^{1/3}.
\ee

At next order, we pose $u = c_1 A^{1/3} + c_2 A^{2/3} + \ldots$. Inserting this in (\ref{tstar}) and keeping again the dominant term in $A$ uniquely determines $c_2$ in terms of $c_1$. Proceeding in this way, one sees that $u$ may be written as an infinite expansion in integer powers of $A^{1/3}$, of which one can compute as many terms as needed. In the present work, we have computed one more term.

The final step is to use the truncated series $t^* = 1 - c_1 A^{1/3} - c_2 A^{2/3} - c_3 A$ to compute $x^*$ at the same order. To do this, we simply use the relation $x^* = t^* + A\, \tli_{-1}(t^*)$ which we again expand in powers of $A^{1/3}$. The result is the one given in (\ref{xstar1d}) {\it without} the exponential factor in the last, linear term. The exact values of the coefficients read
\begin{subequations}
\bea
a_1 &=& \frac{\sqrt{161}-3}{2^{2/3}(\sqrt{161}+1)^{2/3}} \simeq 1.0666, \\
\noalign{\medskip}
a_2 &=& \frac{22\,493 \sqrt{161} - 18\,515}{1\,394\,904} \, a_1^2 \simeq 0.2177,\\
\noalign{\medskip}
a_3 &=& \frac{57\,643 \sqrt{161} + 11\,068\,267}{124\,420\,800} \simeq 0.0948.\phantom{12345}
\eea
\end{subequations}

In the 3D case, the same method can be used with similar results. One finds a series of the form $t^* = 1 - d_1 A^{3/7} - d_2 A^{6/7} - d_3 A^{9/7}$ (next order is $A^{10/7}$) with different (and more complicated) coefficients. For instance, the first coefficient reads
\be
d_1 = \Big[\frac{(661+221\sqrt{881})\, \Gamma[\frac73]^3}{2000}\Big]^{1/7}. 
\ee
The substitution into $x^* = t^* + A\, \tli_{-1/3}(t^*)$ leads to (\ref{xstar3d}), again without the exponential factor, with the following values of the coefficients,
\begin{subequations}
\bea
b_1 &=& {{\frac{(\sqrt{881}-14) \, \Gamma[\frac73]^{3/7}}{2^{9/7} \times 5^{3/7} \, (97\,681+441\sqrt{881})^{1/7}}}} \simeq 0.6617,\phantom{12345}\\
\noalign{\medskip}
b_2 &=& {\frac{4(59\,778\,493 + 2\,788\,893 \sqrt{881})}{578\,742\,115}} \, b_1^2 \simeq 0.4314,\phantom{12345678}\\
\noalign{\medskip}
b_3 &=& -\zeta(-{\textstyle\frac13}) \simeq 0.2773.
\eea
\end{subequations}

Let us finally see the reason for the inclusion of the exponentials in  (\ref{xstar1d}) and (\ref{xstar3d}). All functions $\tli(t)$ have an explicit dependence in $n$ through the subtraction term, see (\ref{tli}). This term is perfectly regular for all values of $t$ and can be expanded around $t=1$, with coefficients explicitly depending on $n$. For the calculation of $t^*$, it turns out that the dependence in $n$ only appears at order $A^{4/3}$ or $A^{10/7}$, and higher. So in each case, 1D or 3D, the first three coefficients $c_1,c_2,c_3$ have no dependence in $n$ at all. 

However when one computes $x^*$ from
\be
x^* = t^* + A \, \li_{-\alpha}(t^*) - A \sum_{j=1}^{n-1} \, j^\alpha t^{*j},
\label{xtstar}
\ee
we do get a dependence in $n$ at order $A$ (and higher), coming from the last term,
\be
\sum_{j=1}^{n-1} \, j^\alpha t^{*j} = \sum_{j=1}^{n-1} \, j^\alpha - (1-t^*) \sum_{j=1}^{n-1} \, j^{\alpha+1} + \ldots
\label{harm}
\ee
If this series is clearly finite in the variable $1-t^*$, $1-t^*$ itself is an infinite series in powers of $A$ so that (\ref{harm}) leads to an infinite series in $A$. Moreover, the coefficients involve the sums $H_{n-1}^{(-\alpha-k)} \equiv \sum_{j=1}^{n-1} j^{\alpha+k} \sim n^{\alpha+k+1}$, $k=0,1,2,\ldots$, which get larger and larger with $k$ (and $n$). In other words, (\ref{xtstar}) and (\ref{harm}) yields for $x^*$ an infinite series in powers of $A$ whose coefficients are actually growing, due to their $n$-dependent part. In such a situation, truncating the series may be problematic.

The trouble comes from the expansion (\ref{harm}). If the limit of the sum when $t^*$ tends to 1 is indeed given by the first term, one can show that the convergence to this limit is exponential in $1-t^*$, something that the truncated series fails to reveal. It is therefore much better to replace it by
\be
\sum_{j=1}^{n-1} \, j^\alpha t^{*j} \simeq H^{(-\alpha)}_{n-1} \, \exp{\big\{\!-\!H^{(-\alpha-1)}_{n-1} (1-t^*)/H^{(-\alpha)}_{n-1}\big\}}.
\ee
which correctly reproduces the first two terms in (\ref{harm}). This improvement does not make much difference if $1-t^*$ is very small and if $n$ is not too large, but may become noticeable for larger values.

Thus the explicit $n$-dependence of $x^*$ in the term linear in $A$ should be corrected to include the exponential decay, and replaced, at dominant order, by
\bea
&& \hspace{-9mm} \frac{n(n-1)}{2} \exp\Big\{\!-\!\frac{(2n-1)}{3} \, c_1 \, F_n^{1/3}\Big\}, \quad \text{(1D case)}\\
&&  \hspace{-9mm} H^{(-1/3)}_{n-1} \, \exp{\Big\{\!-\!\frac{H^{(-4/3)}_{n-1}}{H^{(-1/3)}_{n-1}}\, d_1\, G_n^{3/7}\Big\}}, \quad \text{(3D case)}
\eea
where we have used the well-known formulas $H_{n-1}^{(-1)} = \frac{n(n-1)}2$ and $H_{n-1}^{(-2)} = \frac{n(n-1)(2n-1)}6$. No exact formulas exist for $H_{n-1}^{(-\alpha)}$ when $\alpha$ is not a negative integer. For typical values of $n=4$ and $F_4=G_4=0.01$, the exponential factors equal $0.4689$ and $0.6830$ respectively, and therefore represent a significant correction.


\vspace{7mm}
\hrule

\end{multicols}

\begin{thebibliography}{99}
\bibitem{curv}  Intuitively, the curvature is measured by the change of slope (the amount of ``turning'') divided by the distance travelled along the curve. If the curve is a circle, the curvature is constant, equal to the inverse of the radius of the circle.
\end{thebibliography}
\end{document}